\documentclass[]{article}

\usepackage{graphicx}
\usepackage[capposition=top]{floatrow}
\usepackage[margin=1.in]{geometry}
\usepackage{amsmath}
\usepackage{float}
\usepackage{hyperref}
\usepackage{multicol,lipsum}
\usepackage{algorithm}
\usepackage{algpseudocode}

\title{Bridged treatment comparisons: an illustrative application in HIV treatment}
\author{Paul N Zivich\textsuperscript{1,2}, Stephen R Cole\textsuperscript{1}, Jessie K Edwards\textsuperscript{1}, Bonnie E Shook-Sa\textsuperscript{3}, \\ Alexander Breskin\textsuperscript{4}, Michael G Hudgens\textsuperscript{3}}
\date{%
	\small
	\textsuperscript{1}Department of Epidemiology, Gillings School of Global Public Health, University of North Carolina at Chapel Hill, Chapel Hill, NC\\%
	\textsuperscript{2}Institute of Global Health and Infectious Diseases, University of North Carolina at Chapel Hill, Chapel Hill, NC\\%
	\textsuperscript{3}Department of Biostatistics, Gillings School of Global Public Health, University of North Carolina at Chapel Hill, Chapel Hill, NC\\%
	\textsuperscript{4}Regeneron Pharmaceuticals, Tarrytown, NY\\[2ex]%
	\today
}

\begin{document}

\maketitle

\begin{abstract}
	Comparisons of treatments, interventions, or exposures are of central interest in epidemiology, but direct comparisons are not always possible due to practical or ethical reasons. Here, we detail a fusion approach to compare treatments across studies. The motivating example entails comparing the risk of the composite outcome of death, AIDS, or greater than a 50\% CD4 cell count decline in people with HIV when assigned triple versus mono antiretroviral therapy, using data from the AIDS Clinical Trial Group (ACTG) 175 (mono versus dual therapy) and ACTG 320 (dual versus triple therapy). We review a set of identification assumptions and estimate the risk difference using an inverse probability weighting estimator that leverages the shared trial arms (dual therapy). A fusion diagnostic based on comparing the shared arms is proposed that may indicate violation of the identification assumptions. Application of the data fusion estimator and diagnostic to the ACTG trials indicates triple therapy results in a reduction in risk compared to monotherapy in individuals with baseline CD4 counts between 50 and 300 cells/mm\textsuperscript{3}. Bridged treatment comparisons address questions that none of the constituent data sources could address alone, but valid fusion-based inference requires careful consideration of the underlying assumptions.
\end{abstract}

\section*{Introduction}
Comparisons between new and existing treatments are of central interest in epidemiology and medical science. However, direct comparisons between two treatments is often infeasible due to practical or ethical reasons. For example, once an efficacious treatment is developed, it is typically no longer ethical to conduct placebo-controlled trials, thus precluding direct comparisons between new treatments and placebo. Instead, informal transitivity arguments are often made to compare a set of studies through shared intermediary treatments \cite{Catala2014, Lumley2002, Rouse2017}. For example, if treatment A is better than B, and B is better than C, then A is better than C. However, such simple comparisons do not provide a quantitative estimate and may even be misleading for a variety of reasons (e.g., studies may sample from different populations, define endpoints differently, or have different rates of loss to follow-up). The exact conditions for valid comparisons of treatments across studies are often left implicit and not assessed.

As a motivating example, consider different combinations of antiretroviral therapies for treatment of HIV infection. The AIDS Clinical Trials Group (ACTG) 175 randomized trial found dual therapy to be superior to monotherapy \cite{Hammer1996actg175}, and the ACTG 320 randomized trial found triple therapy to be superior to dual therapy \cite{Hammer1997actg320}. At the time of those trials, researchers, agencies, patients, and other stakeholders might be interested in comparing the efficacy of triple therapy relative to monotherapy. Simple comparisons across trials reflect the efficacy of triple versus monotherapy \textit{and} underlying differences between the trials, and thus may be misleading. As an example, suppose the prevalence of drug resistance increased in the intervening period between these trials so that dual therapy became less effective for treatment of HIV infection between trials. Na\"ively comparing the triple and monotherapy arms across trials could then result in a biased estimate.

Bridged treatment comparisons, a particular fusion study design \cite{Cole2022}, have been proposed as an alternative approach to compare treatments across trials \cite{Breskin2021}.  Here, a detailed step-by-step guide is given with the ACTG 175 and ACTG 320 trials serving as the motivating example. A sufficient set of assumptions under which bridged comparisons are identified are provided and inverse probability weighting estimators that can account for differences between the trial populations analytically are described. Corresponding data and code in SAS, R, and Python are provided. Finally, graphical and statistical diagnostics are proposed of the underlying assumptions for the comparison.

\section*{Trials}

ACTG 175 was a double-masked randomized trial comparing four antiretroviral therapy regimens among HIV-1 infected participants with CD4 counts between 200 and 500 cells/mm\textsuperscript{3} between December 1991 and November 1994 \cite{Hammer1996actg175}. Trial arms consisted of zidovudine-only, zidovudine-didanosine, zidovudine-zalcitabine, and didanosine-only. A total of 2493 participants were recruited from sites in the United States. Eligible participants were at least 12 years of age, had no history of previous AIDS-defining illness (expect minimal Kaposi's sarcoma), and had a Karnofsky score of at least 70. Here, we restricted to three of the trial arms: zidovudine-didanosine or zidovudine-zalcitabine (combined as dual therapy), and zidovudine (mono therapy). The composite outcome was time from randomization to AIDS, death, or a 50\%+ decline in CD4 cell count from baseline.

ACTG 320 was a double-masked randomized trial comparing dual nucleoside antiretroviral therapy with a protease inhibitor (zidovudine-lamivudine-indinavir) to dual nucleoside antiretroviral therapy (zidovudine-lamivudine) among HIV-1 infected participants with CD4 cell counts below 200 cells/mm\textsuperscript{3} between January 1996 to February 1997 \cite{Hammer1997actg320}.  A total of 1156 participants were recruited from sites in the United States. Eligible participants were at least 16 years old, had a Karnofsky score of at least 70, had at least 3 months of prior treatment with zidovudine, no more than one week of prior treatment with lamivudine, and no prior treatment with protease inhibitors. The outcome was time from randomization to AIDS or death.

To harmonize data from these trials, we restricted the ACTG 175 to participants 16 years or older with at least three months of prior zidovudine treatment and no prior non-zidovudine antiretroviral therapy (813 of 2493 participants), administratively censored all participants at one year, and used the composite outcome of AIDS, death, or a 50\%+ decline in CD4 cell count. Both trials collected a variety of baseline characteristics (Table 1). Note the non-overlapping baseline CD4 cell counts due to differing eligibility criteria between trials (Figure A1), a key point returned to below.

\section*{Bridged Comparison Design and Estimation}

The fusion study design presented here consists of five steps: (1) definition of the target population, (2) definition of the parameter of interest, (3) identification assumptions and selection of an appropriate estimator, (4) diagnostic of the fusion, and (5) estimation.

The following notation will be used. Let $T_i^a$ indicate the potential event time for individual $i$ when assigned treatment $a$, where $a=1$ is mono therapy, $a=2$ is dual therapy, and $a=3$ is triple therapy. Let $A_i = \{1,2,3\}$ indicate the assigned antiretroviral therapy, and $T_i = \sum_{a\in\{1,2,3\}} T_i^a I(A_i=a)$ indicate the event time corresponding to the assigned treatment, where $I(B)=1$ if $B$ is true and $I(B)=0$ otherwise. The observed event times may be right censored due to drop-out or the end of follow-up. Let $C_i$ indicate the censoring time such that the observed event times are $T_i^* = \text{min}(T_i, C_i)$, and $\delta_i = I(T_i^* = T_i)$. Finally, let $W_i$ indicate a vector of observed baseline covariates measured in both trials, $V_i$ indicate a separate vector of observed baseline covariates (the distinction between $W_i$ and $V_i$ is made below), and $S_i=1$ if individual $i$ participated in ACTG 320 and $S_i=0$ indicate participation in ACTG 175.  Hereafter, we assume the observed data consists of $n$ independent observations of $(S_i, V_i, W_i, A_i, T_i^*, \delta_i)$.

\subsection*{Step 1: Definition of the target population}

Due to possible treatment effect heterogeneity across trials, the target population must be clearly defined. While trial participants are not typically selected via a formal sampling design, a trial-specific population can still be defined as the set of individuals who meet the trial's inclusion/exclusion criteria and would self-select to participate in the trial. Therefore, the target parameter can be defined for the ACTG 175 population, the ACTG 320 population, the combined population of both trials, or some external population (which would require additional data from or assumptions about that external population). Here, we chose to estimate the parameter in the ACTG 320 population.

\subsection*{Step 2: Definition of the parameter of interest}

The motivating quantity of interest is the risk difference at one-year of follow-up comparing the risk of the composite outcome had everyone been assigned triple therapy with the risk of the composite outcome had everyone been assigned monotherapy in the ACTG 320 population. When there is non-adherence, either the intent-to-treat or per-protocol parameter may be of interest. The intent-to-treat parameter compares outcomes following treatment assignment, regardless of treatment received, whereas the per-protocol parameter quantifies the effect of treatment when there is perfect adherence to assigned treatment \cite{Rudolph2020}. Here, focus was on the intent-to-treat parameter, $RD^{3-1}(t)$, which can be expressed as
\begin{equation}
	\label{Eq1}
	\begin{split}
		RD^{3-1}(t) = & \Pr(T^{a=3} < t | S=1) - \Pr(T^{a=3} < t | S=1) \\
		            = & R^3(t) - R^1(t)
	\end{split}
\end{equation}
This parameter corresponds to a trial conducted in the ACTG 320 trial population but where participants were randomized to either triple or monotherapy. Since no individuals in ACTG 320 were randomized to monotherapy, $R^1(t)$ is not identified from the ACTG 320 data. Therefore, we re-express $RD^{3-1}(t)$ in terms of quantities that may be possible to identify using data from both trials. With the risk of the outcome at time $t$ under dual therapy acting as a `bridge', the parameter can instead be written as 
\begin{equation}
	\label{Eq2}
	\begin{split}
		RD^{3-1}(t) = & \left(R^3(t) - R^2(t)\right) - \left(R^2(t) - R^1(t)\right) \\
		= & RD^{3-2}(t) - RD^{2-1}(t)
	\end{split}
\end{equation}~

\subsection*{Step 3: Identification assumptions and target parameter estimator}

In this section, a set of sufficient identification assumptions are provided with corresponding parameter estimators.

\subsubsection*{Step 3a: Identification assumptions and estimation of triple versus dual therapy}

The ACTG 320 trial was designed specifically to estimate $RD^{3-2}(t)$, so it is natural to use data from this trial to estimate this risk difference. As triple and dual antiretroviral therapy regimens were randomly assigned in the ACTG 320 trial, the identification assumptions of exchangeability (i.e., assigned antiretroviral therapy and potential outcomes were independent) \cite{Hernan2006}, and positivity (i.e., all individuals had non-zero probability of assignment to each antiretroviral therapy regimen) \cite{Westreich2010, Zivich2022} are given by design. The exchangeability and positivity assumptions are expressed, respectively, as
\[\Pr(T^a < t | S=1) = \Pr(T^a < t | A=a, S=1) \text{ for } a\in\{2,3\}\]
\[\Pr(A=a | S=1) > 0 \text{ for } a\in\{2,3\}\]
Under these identification assumptions and causal consistency \cite{Cole2009}, $RD^{3-2}(t)$ can be expressed in terms of the observed data distribution in the absence of censoring \cite{Breskin2021, Robins2008}. However, some participants were right censored prior to one-year and hence the event time was not observed for all participants. Additionally, censoring due to loss to follow-up may be driven by both baseline characteristics of participants and assigned treatment. Therefore, exchangeability between censored and uncensored participants conditional on $W$ and $A$, and positivity is assumed at each time point:
\[\Pr(T < t | A,W,S=1) = \Pr(T < t | C,A,W,S=1)\]
\[\Pr(C > T | A=a,W=w,S=1) > 0 \text{ for all } a,w \text{ such that } \Pr(A=a,W=w | S=1) > 0\]
To estimate $RD^{3-2}(t)$, we use the following inverse probability weighted estimator:
\[\widehat{RD}_{320}^{3-2}(t) = \widehat{R}_{320}^{3}(t) - \widehat{R}_{320}^{2}(t)\]
where
\begin{equation}
	\label{Eq3}
	\begin{split}
		\widehat{R}_{320}^{a}(t) = \frac{1}{n_{320}} \left(\sum_{i=1}^{n} \frac{I(A_i = a) \; I(S_i =1) \; I(T_i^* \le t) \delta_i}{\Pr(A_i=a | S_i=1) \pi_C(T_i^*,W_i,A_i,S_i,\hat{\alpha})}\right) \text{ for } a\in\{2,3\}
	\end{split}
\end{equation}
the subscript 320 emphasizes that the ACTG 320 trial data is being used for estimation, and $n_{320} = \sum_{i=1}^{n} I(S_i = 1)$ is the number of participants in the ACTG 320 trial. The numerator of equation \ref{Eq3} corresponds to whether an event was observed by time $t$ (i.e., $I(T_i^* \le t) \delta_i$) among individuals in the ACTG 320 (i.e., $I(S_i=1)$) in the assigned treatment group (i.e., $I(A_i=a)$ for $a\in\{2,3\}$). The denominator of equation \ref{Eq3} consists of the randomization probability of the corresponding treatment (i.e., $\Pr(A_i = a | S_i = 1)$ for $a\in\{2,3\}$) and a model-estimated conditional probability of remaining uncensored (i.e., $\pi_C(T_i^*,W_i,A_i,S_i;\hat{\alpha})$) described further below. The denominator effectively reweights the observed events to account for those who were censored and were in the other treatment arm.

In this analysis, $W$ consisted of gender, race, injection drug use, age, and Karnofsky score categories. The nuisance parameters, $\alpha$, in $\pi_C(T_i^*,W_i,A_i,S_i,\hat{\alpha})$ of \ref{Eq3} were estimated by fitting a Cox model of time until drop-out via maximum partial likelihood \cite{Cox1972}. To estimate the probability of remaining uncensored at $T_i^*$, the cumulative hazard function was nonparametrically estimated using the Breslow estimator \cite{Breslow1972, Lin2007}. For flexibility in the Cox model, we stratified by antiretroviral therapy and age was modeled using a restricted quadratic spline with knots placed at the 5\textsuperscript{th}, 35\textsuperscript{th}, 65\textsuperscript{th}, and 95\textsuperscript{th} percentiles. While the treatment probabilities were known by design, $\Pr(A_i=a | S_i=1)$ was instead estimated via an intercept-only logistic model, as estimation is expected to lead to a smaller asymptotic variance \cite{vdL2003}. One could also include covariates predictive of the outcome in this model for the purpose of increasing precision \cite{Williamson2014, Morris2022}.

\subsubsection*{Step 3b: Identification assumptions and estimation of dual versus monotherapy}

Conditional on $S=0$, exchangeability and positivity for dual and monotherapy are given by design in the ACTG 175 trial:
\[\Pr(T^a < t | S=0) = \Pr(T^a < t | A=a, S=0) \text{ for } a\in\{1,2\}\]
\[\Pr(A=a | S=0) > 0 \text{ for } a\in\{1,2\}\]
Conditional exchangeability and positivity for censoring are also assumed:
\[\Pr(T < t | A,W,S=0) = \Pr(T < t | C,A,W,S=0)\]
\[\Pr(C > T | A=a,W=w,S=0) > 0 \text{ for all } a,w \text{ such that } \Pr(A=a,W=w | S=0) > 0\]
While the covariates included in $W$ are the same for both trials, this need not be the case.

The ACTG 175 trial was designed to estimate $\Pr(T^a < t | S=0)$ for $a\in\{1,2\}$, which need not equal $\Pr(T^a < t | S=1)$ (i.e., we are unwilling to assume both trials are random samples of the same population). Therefore, we need to account for differences between populations. To do so, the estimator in equation \ref{Eq3} is extended with developments from the generalizability and transportability literature \cite{Cole2010, Lesko2017, Westreich2017}. This extension is done under the assumption that selection into study populations is exchangeable conditional on a set of baseline covariates $V$, common to the two trials. Conditional exchangeability and positivity of sampling are assumed:
\[\Pr(T^a < t | V,S=0) = \Pr(T^a < t | V,S=1) \text{ for } a\in\{1,2\}\]
\[\Pr(S=0 | V=v) > 0 \text{ for all } v \text{ such that } \Pr(S=1 | V=v) > 0\]
Here, $V$ includes demographics, risk factors, and other contextual factors that both differ between trial populations and are related to the composite outcome. 

The estimator in equation \ref{Eq3} is extended to account for differences between trial populations via inverse odds of sampling weights \cite{Westreich2017}. The inverse odds weights are defined as
\[\frac{1 - \Pr(S_i = 0 | V_i)}{\Pr(S_i = 0 | V_i)}\]
where $\Pr(S_i = 0 | V_i)$ is the probability of being a member of the ACTG 175 sample conditional on the baseline covariates $V$. The conditional probability of being in ACTG 175 is estimated using a model (e.g., logistic regression), where $\pi_S(V_i, \hat{\beta})$ represents a model-based estimate of $\Pr(S_i = 0 | V_i)$. Here, the baseline covariates included in the censoring and sampling nuisance models are the same (i.e., $W=V$) but this need not be the case.

The inverse probability weights estimator for $RD^{2-1}(t)$, which includes the inverse odds of sampling weights, is
\[\widehat{RD}_{175}^{2-1}(t) = \widehat{R}_{175}^{2}(t) - \widehat{R}_{175}^{1}(t)\]
where
\begin{equation}
	\label{Eq4}
	\widehat{R}_{175}^{2-1}(t) = \frac{1}{\hat{n}_{175}} \left( \sum_{i=1}^{n} 	\frac{I(A_i = a) \; I(S_i =1) \; I(T_i^* \le t) \delta_i}{\Pr(A_i=a | S_i=1) \pi_C(T_i^*,W_i,A_i,S_i,\hat{\alpha})}
	\times 
	\frac{1 - \pi_S(V_i, \hat{\beta})}{\pi_S(V_i, \hat{\beta})}
	\right)  \text{ for } a\in\{1,2\}
\end{equation}
and $\hat{n}_{175} = \sum_{i=1}^{n} I(S_i = 0) \frac{1 - \pi_S(V_i, \hat{\beta})}{\pi_S(V_i, \hat{\beta})}$ is the inverse odds of sampling weighted sample size. If the odds weight model is correctly specified, $\hat{n}_{175}$ should be approximately equal to $n_{320}$. To estimate $\pi_S(V_i, \hat{\beta})$, data from both trials were stacked together. The conditional probability of remaining uncensored was estimated via a Cox model as detailed in Step 3a. Lastly, $\Pr(A_i=a | S_i=0)$ was estimated via an intercept-only logistic model.

\subsection*{Step 4: Diagnostic for the fusion}

To assess the validity of the bridged comparison, we propose graphical and statistical diagnostics. Because $\widehat{R}_{320}^{2}(t)$ and $\widehat{R}_{175}^{2}(t)$ are estimators for the same quantity (i.e., risk under dual therapy for the ACTG 320 population), the difference, $\widehat{R}_{320}^{2}(t) - \widehat{R}_{175}^{2}(t)$, should be close to zero at all $t$ when the previously described assumptions hold, including correct model specification. Therefore, we propose comparing the estimated risk between the shared arms across trials. To assess potential non-zero differences between $\widehat{R}_{320}^{2}(t)$ and $\widehat{R}_{175}^{2}(t)$, graphical or statistical comparisons can be made. 

For graphical comparisons, tiwster plots of the differences in estimated risks between the shared arms can be used \cite{Zivich2021}. These plots enable visual assessment of whether any differences occur over the follow-up period. As a statistical comparison, a test statistic based on the integrated risk difference between the shared arms is used \cite{Pepe1989, Pepe1991, Cole2009a}. As the risk functions are expected to be equal over the interval $[0, \tau)$, the expected integrated risk difference is zero under the previous assumptions \cite{Pepe1989}. To compute a Wald-type test, the standard error of the integrated risk difference is estimated using a nonparametric bootstrap \cite{Kulesa2015}, with an algorithm provided in Appendix 1. In this context, small P values are suggestive of at least one of the previous assumptions being violated.

In addition to the bridging diagnostic described above, imbalance in covariates $V_i$ in the reweighted $S_i=0$ data can also be assessed \cite{Austin2015}. However, unlike checking for imbalances in the observed covariates, the proposed diagnostics may detect the present of unmeasured covariates predictive of trial participation and the outcome.

\subsection*{Step 5: Estimation}

The risk difference for triple versus monotherapy at time $t$ was estimated by
\[\widehat{RD}^{3-1}(t) = \widehat{RD}_{320}^{3-2}(t) + \widehat{RD}_{175}^{2-1}(t)\]
To estimate the variance, a nonparametric bootstrap was used \cite{Kulesa2015}. Briefly, $n_1$ and $n_0$ observations were separately resampled with replacement from each trial, then the nuisance models and risk difference function were estimated using those resampled data sets. The standard error of $\widehat{RD}^{3-1}(t)$ was estimated by the standard deviation of the resampled risk difference estimates. Corresponding Wald-type 95\% confidence intervals (CI) were constructed via $\widehat{RD}^{3-1}(t) \pm z_{0.975} \hat{\sigma}(t)$, where $z_{0.975}$ is the 97.5\textsuperscript{th} quantile of a standard normal distribution and $\hat{\sigma}(t)$ is the bootstrap estimated standard error.

\section*{Results}

\subsection*{Simulation Results}

To assess the performance of the diagnostic test, a brief simulation study was conducted. Details on data generation are provided in Appendix 2. Type 1 error and power were compared for difference significance level thresholds of 0.05, 0.10, and 0.20. The type 1 error was estimated by the proportion of P values below the threshold under the correctly specified nuisance models. Power was estimated by the proportion of P values below the specified significance level when a covariate was excluded from the sampling model. The type 1 error was near the nominal level (Appendix Table A1) and power was 1 despite the magnitude of bias being small relative to the standard error of the estimator. CIs covered at the nominal level when the sampling model was correctly specified, with slight increases in coverage when restricting the simulated data sets to where the diagnostic P value was above the significance threshold (Appendix Table A2-A3).

\subsection*{Application}

Before comparing triple to monotherapy, the proposed graphical and statistical diagnostics were applied. First, a sizable difference was observed the estimated risk functions for the composite outcome when the ACTG 175 dual arm was not reweighted to the ACTG 320 population (Figure 1A, $P < 0.001$). After the ACTG 175 was re-weighted using the inverse odds of sampling weights, the difference between the risks of the dual therapy arms was reduced but not ameliorated (Figure 1B, $P < 0.001$). As CD4 count is predictive of the composite outcome and was not included in the sampling model as in Breskin et al. (2021) \cite{Breskin2021}, differences in baseline CD4 likely explain why the ACTG 175 dual arm (where baseline CD4 counts are higher) appears beneficial. Addition of CD4 count to the sampling model did not correct for this discrepancy (Figure 1C, $P < 0.001$) and $\hat{n}_{175}$ was nearly double $n_{320}$, likely due to deterministic nonpositivity of $S=0$ for lower values of CD4. Restricting analyses to where there was overlap of CD4 counts between trials (50-300 cells/mm\textsuperscript{3}) reduced the differences between the dual therapy arms (Figure 1D, $P = 0.09$). Therefore, the analysis proceeded with the CD4 restricted population (Table 2). Notice that the restriction by CD4 alters the target population.

The estimated risk difference function for the composite outcome under triple versus monotherapy is depicted in Figure 2. These results indicate that assignment to triple therapy is preferred over mono therapy for the one-year risk of AIDS, death, or 50\%+ decline in CD4. At one-year of follow-up the risk of the composite outcome was 21 percentage points lower (95\% CI: -33\%, -8\%) when assigned triple therapy compared to mono therapy in the ACTG 320 population with CD4 counts between 50-300 cells/mm\textsuperscript{3} at baseline.

\section*{Discussion}

This paper provides a step-by-step guide to bridged treatment comparisons in the context of HIV and antiretroviral therapy. The described fusion estimator accounts for observed differences between trial populations and informative censoring through inverse probability weighting, comparing treatments across trials through a shared intermediate arm. To evaluate the fusion, we proposed diagnostic procedures that compare the estimated risks of the shared arms. We found that assignment to triple therapy was preferred to monotherapy in terms of preventing AIDS, death, or a 50\%+ decline in CD4 over a one-year period. However, the target population had to be restricted due to deterministic non-positivity between the two trials.

Bridged comparisons between treatment may be necessary in some settings, where ethical or feasibility considerations rule out the use of a direct comparison. Other approaches for comparisons of treatments across trials include network meta-analysis \cite{Lumley2002,Rouse2017}, placebo arms of external trials \cite{Donnell2023}, matching adjusted indirect comparisons \cite{Signorovitch2012}, use of cross-sectional data and recency tests to stand-in for placebo or standard of care arms \cite{Gao2021, Gao2022}, and use of priors on adherence-efficacy relationships to model a placebo arm for non-inferiority trials \cite{Glidden2020, Glidden2021}. Importantly, bridged treatment comparisons offer an intuitive diagnostic for fusing trials together. However, bridged treatment comparisons require the use of individual-level data from both trials with overlapping covariates.

Here, the intent-to-treat parameter was estimated, but bridged treatment comparisons can also be used to estimate per-protocol parameters \cite{Breskin2021}. However, treatment exchangeability is no longer given by design for per-protocol parameters. The intent-to-treat parameter may also be of greater practical relevance because the per-protocol parameter quantifies treatment effects when there is perfect adherence, whereas in most real-world settings adherence is often less than perfect. However, some assumption regarding adherence across trials becomes necessary, unlike intent-to-treat analyses of a single trial. Specifically, the set of covariates for the sampling exchangebility assumption must also include contextual factors that differed between populations and were related to adherence to assigned treatment. If adherence was directly affected by an unmeasured covariate and the distribution of that covariate differed between trial populations, then sampling exchangeability may not hold. This concern regarding adherence arises in general when transporting intent-to-treat parameters across populations \cite{Westreich2015, Dahabreh2022}.

In the motivating example, the proposed diagnostic indicated that na\"ively fusing data from the two ACTG trials may not be valid. While the differences in eligibility by CD4 count are apparent in the inclusion criteria, data sources from different places or times may differ in ways not readily apparent in inclusion or exclusion criteria. The proposed diagnostics can be used to assess whether at least one of the assumptions (i.e., identification assumptions or specification of nuisance models) does not hold. It should be noted, however, that the lack of a difference between the shared arms is not sufficient to conclude that all assumptions hold true. As a counterexample, consider when the censoring models for the shared arms are correctly specified but the censoring model for the non-overlapping treatment arms is incorrectly specified. In this case, the proposed diagnostics will fail to detect the model misspecification. Alternatively, if multiple variables required for sampling exchangeability were omitted, the impact of those missing variables could cancel or mask each other such that no difference between the shared arms is detected. These counterexamples are akin to limitations of other diagnostics, such as comparison of the g-formula under the natural course to the observed data \cite{Keil2014}, checking for imbalances in the pseudo-population when using inverse probability weighting \cite{Austin2015},  or testable implications for generalizability and meta-analyses \cite{Stuart2011,Dahabreh2020}.

There are several considerations when applying and interpreting results from the diagnostic test. First, small P values may occur due to sampling variability. Forgoing fusion of trials in these cases, where results from a bridged treatment comparison would be valid otherwise, means that the proposed diagnostic can be conservative when deciding whether to bridge trials. Further, small P values may also result when fusing data from large trials even if there are only minor violations in the assumptions. Likewise, large P values should also be interpreted with care, especially when sample sizes are small, and do not necessarily imply that all assumptions hold. With high loss-to-follow-up, these concerns are exacerbated due to instability in the differences at times closer to $\tau$ \cite{Pepe1989}. Down-weighting differences at later times, as done with the weighted Kaplan-Meier test statistic \cite{Pepe1989, Pepe1991}, could help to ameliorate this issue. Finally, positivity differences in the risk functions may cancel out negative differences when the risk functions cross each other. Therefore, the validity of bridged treatment comparisons should also be based on the graphical comparison and background knowledge. In addition, inference is complicated when diagnostics lead to revised analyses. The randomness of data-driven decisions based on the diagnostic is not accounted for by the bootstrap standard error estimator employed in this paper. While this randomness could be accounted for by extending the bootstrap procedure to include analytic decision points \cite{Harrell1996}, such an approach requires a well-defined, pre-specified decision process. 

Bridged treatment comparisons and the proposed diagnostic test are similar to test-then-pool designs \cite{Viele2014, Li2020}, where external or historical control data is pooled with the control arm of a trial if the control arms are deemed similar via statistical test. However, there are two important differences. First, the statistical test in test-then-pool designs typically compare binary or continuous endpoints measured at a single time \cite{Viele2014, Li2020}. The proposed diagnostic test compares the entire risk functions. Second, test-then-pool analyses can still address the motivating question when the trial control and historical control data are deemed to differ by dropping the historical data from the analysis. On the other hand, bridged treatment comparisons cannot be made without pooling trial data as both trials are needed to address the motivating question.

There are several possible areas of future methodological developments related to bridged treatment comparisons. First, bridged comparisons could further incorporate time-varying covariates predictive of drop-out and the outcome. In the context of the HIV example, the presence of side-effects of protease inhibitors may drive differential drop-out \cite{Carr1998}. Second, methods for bridged comparisons could be developed that rely on weaker or alternative sets of assumptions \cite{ShookSa2023, Zivich2023p}. Third, more efficient estimators relative to the described inverse probability weighting estimator used here may be developed by utilizing an outcome model (i.e., an augmented inverse probability weighting estimator) \cite{Daniel2014,Rotnitsky2006}. Fourth, bounded estimators could also be considered. The inverse probability weighting estimators are not bounded to be less than one. Estimated risks above one may result from extreme weights, which is also indicative of violations of assumptions, such as positivity. Fifth, further research into the theoretical properties of the permutation test is warranted. Lastly, the estimation approach relied on parametric or semiparametric models to estimate the corresponding nuisance parameters. While restricted quadratic splines were used to model age and censoring models were estimated separately for trials, such modeling choices may not be sufficiently flexible to meet the correct model specification assumption generally. Instead, more flexible data-adaptive or machine learning algorithms could be considered to estimate nuisance parameters \cite{Zivich2021a, Naimi2017}.

\section*{Acknowledgments}
This work was supported in part by R01-AI157758 (PNZ, SRC, BES, JKE), P30-AI050410 (SRC, MGH, BES), K01AI125087 (JKE), and T32-AI007001 (PNZ). The authors would like to thank the HIV/STI trainees at the University of North Carolina at Chapel Hill for their feedback and suggestions. Corresponding code is available at https://github.com/pzivich/publications-code

\small
\bibliography{biblio}{}

\begin{thebibliography}{10}

\bibitem{Catala2014}
F.~Catal{\'a}-L{\'o}pez, B.~Hutton, and D.~Moher, ``The transitive property
  across randomized controlled trials: if {B} is better than {A}, and {C} is
  better than {B}, will {C} be better than {A}?,'' {\em Revista {E}spanola de
  {C}ardiologia (English ed.)}, vol.~67, no.~8, pp.~597--602, 2014.

\bibitem{Lumley2002}
T.~Lumley, ``Network meta-analysis for indirect treatment comparisons,'' {\em
  Statistics in {M}edicine}, vol.~21, no.~16, pp.~2313--2324, 2002.

\bibitem{Rouse2017}
B.~Rouse, A.~Chaimani, and T.~Li, ``Network meta-analysis: an introduction for
  clinicians,'' {\em Internal and {E}mergency {M}edicine}, vol.~12, no.~1,
  pp.~103--111, 2017.

\bibitem{Hammer1996actg175}
S.~M. Hammer, D.~A. Katzenstein, M.~D. Hughes, H.~Gundacker, R.~T. Schooley,
  R.~H. Haubrich, W.~K. Henry, M.~M. Lederman, J.~P. Phair, M.~Niu, {\em
  et~al.}, ``A trial comparing nucleoside monotherapy with combination therapy
  in {HIV}-infected adults with {CD}4 cell counts from 200 to 500 per cubic
  millimeter,'' {\em New {E}ngland {J}ournal of {M}edicine}, vol.~335, no.~15,
  pp.~1081--1090, 1996.

\bibitem{Hammer1997actg320}
S.~M. Hammer, K.~E. Squires, M.~D. Hughes, J.~M. Grimes, L.~M. Demeter, J.~S.
  Currier, J.~J. Eron~Jr, J.~E. Feinberg, H.~H. Balfour~Jr, L.~R. Deyton, {\em
  et~al.}, ``A controlled trial of two nucleoside analogues plus indinavir in
  persons with human immunodeficiency virus infection and {CD}4 cell counts of
  200 per cubic millimeter or less,'' {\em New {E}ngland {J}ournal of
  {M}edicine}, vol.~337, no.~11, pp.~725--733, 1997.

\bibitem{Cole2022}
S.~R. Cole, J.~K. Edwards, A.~Breskin, S.~Rosin, P.~N. Zivich, B.~E. Shook-Sa,
  and M.~G. Hudgens, ``Illustration of {T}wo {F}usion {D}esigns and
  {E}stimators,'' {\em American {J}ournal of {E}pidemiology}.

\bibitem{Breskin2021}
A.~Breskin, S.~R. Cole, J.~K. Edwards, R.~Brookmeyer, J.~J. Eron, and A.~A.
  Adimora, ``Fusion designs and estimators for treatment effects,'' {\em
  Statistics in {M}edicine}, vol.~40, no.~13, pp.~3124--3137, 2021.

\bibitem{Rudolph2020}
J.~E. Rudolph, A.~I. Naimi, D.~J. Westreich, E.~H. Kennedy, and E.~F.
  Schisterman, ``Defining and identifying per-protocol effects in randomized
  trials,'' {\em Epidemiology}, vol.~31, no.~5, pp.~692--694, 2020.

\bibitem{Hernan2006}
M.~A. Hern{\'a}n and J.~M. Robins, ``Estimating causal effects from
  epidemiological data,'' {\em Journal of {E}pidemiology \& {C}ommunity
  {H}ealth}, vol.~60, no.~7, pp.~578--586, 2006.

\bibitem{Westreich2010}
D.~Westreich and S.~R. Cole, ``Invited commentary: positivity in practice,''
  {\em American {J}ournal of {E}pidemiology}, vol.~171, no.~6, pp.~674--677,
  2010.

\bibitem{Zivich2022}
P.~N. Zivich, S.~R. Cole, and D.~Westreich, ``Positivity: {I}dentifiability and
  {E}stimability,'' {\em ar{X}iv preprint ar{X}iv:2207.05010}, 2022.

\bibitem{Cole2009}
S.~R. Cole and C.~E. Frangakis, ``The consistency statement in causal
  inference: a definition or an assumption?,'' {\em Epidemiology}, vol.~20,
  no.~1, pp.~3--5, 2009.

\bibitem{Robins2008}
J.~M. Robins and M.~A. Hern{\'a}n, ``Estimation of the causal effects of
  time-varying exposures,'' in {\em Longitudinal {D}ata {A}nalysis}
  (G.~Fitzmaurice, M.~Davidian, G.~Verbeke, and G.~Molenberghs, eds.),
  pp.~567--614, Chapman and Hall/CRC, 2008.

\bibitem{Cox1972}
D.~R. Cox, ``Regression {M}odels and {L}ife-{T}ables,'' {\em Journal of the
  {R}oyal {S}tatistical {S}ociety: {S}eries {B} ({M}ethodological)}, vol.~34,
  no.~2, pp.~187--202, 1972.

\bibitem{Breslow1972}
N.~E. Breslow, ``Discussion of the paper by {D}.{R}. {C}ox,'' {\em Journal of
  the {R}oyal {S}tatistical {S}ociety: {S}eries {B} ({M}ethodological)},
  vol.~34, no.~2, pp.~216--217, 1972.

\bibitem{Lin2007}
D.~Lin, ``On the {B}reslow estimator,'' {\em Lifetime {D}ata {A}nalysis},
  vol.~13, no.~4, pp.~471--480, 2007.

\bibitem{vdL2003}
M.~J. Van~der Laan and J.~M. Robins, {\em Unified methods for censored
  longitudinal data and causality}, vol.~5.
\newblock Springer, 2003.

\bibitem{Williamson2014}
E.~J. Williamson, A.~Forbes, and I.~R. White, ``Variance reduction in
  randomised trials by inverse probability weighting using the propensity
  score,'' {\em Statistics in {M}edicine}, vol.~33, no.~5, pp.~721--737, 2014.

\bibitem{Morris2022}
T.~P. Morris, A.~S. Walker, E.~J. Williamson, and I.~R. White, ``Planning a
  method for covariate adjustment in individually randomised trials: a
  practical guide,'' {\em Trials}, vol.~23, no.~1, pp.~1--17, 2022.

\bibitem{Cole2010}
S.~R. Cole and E.~A. Stuart, ``Generalizing evidence from randomized clinical
  trials to target populations: the {ACTG} 320 trial,'' {\em American {J}ournal
  of {E}pidemiology}, vol.~172, no.~1, pp.~107--115, 2010.

\bibitem{Lesko2017}
C.~R. Lesko, A.~L. Buchanan, D.~Westreich, J.~K. Edwards, M.~G. Hudgens, and
  S.~R. Cole, ``Generalizing study results: a potential outcomes perspective,''
  {\em Epidemiology}, vol.~28, no.~4, p.~553, 2017.

\bibitem{Westreich2017}
D.~Westreich, J.~K. Edwards, C.~R. Lesko, E.~Stuart, and S.~R. Cole,
  ``Transportability of trial results using inverse odds of sampling weights,''
  {\em American {J}ournal of {E}pidemiology}, vol.~186, no.~8, pp.~1010--1014,
  2017.

\bibitem{Zivich2021}
P.~N. Zivich, S.~R. Cole, and A.~Breskin, ``Twister {P}lots for
  {T}ime-to-{E}vent {S}tudies,'' {\em American {J}ournal of {E}pidemiology},
  vol.~190, no.~12, pp.~2730--2731, 2021.

\bibitem{Pepe1989}
M.~S. Pepe and T.~R. Fleming, ``Weighted {Kaplan}-{Meier} {Statistics}: {A}
  {Class} of {Distance} {Tests} for {Censored} {Survival} {Data},'' {\em
  Biometrics}, vol.~45, no.~2, pp.~497--507, 1989.
\newblock Publisher: [Wiley, International Biometric Society].

\bibitem{Pepe1991}
M.~S. Pepe and T.~R. Fleming, ``Weighted {Kaplan}-{Meier} {Statistics}: {Large}
  {Sample} and {Optimality} {Considerations},'' {\em Journal of the Royal
  Statistical Society. Series B (Methodological)}, vol.~53, no.~2,
  pp.~341--352, 1991.
\newblock Publisher: [Royal Statistical Society, Wiley].

\bibitem{Cole2009a}
S.~R. Cole, H.~Chu, and L.~Nie, ``Nonparametric estimator of relative time with
  application to the acyclovir prevention trial,'' {\em Clinical {T}rials},
  vol.~6, no.~4, pp.~320--328, 2009.

\bibitem{Kulesa2015}
A.~Kulesa, M.~Krzywinski, P.~Blainey, and N.~Altman, ``Sampling distributions
  and the bootstrap,'' {\em Nature {M}ethods}, vol.~12, pp.~477--478, 2015.

\bibitem{Austin2015}
P.~C. Austin and E.~A. Stuart, ``Moving towards best practice when using
  inverse probability of treatment weighting ({IPTW}) using the propensity
  score to estimate causal treatment effects in observational studies,'' {\em
  Statistics in {M}edicine}, vol.~34, no.~28, pp.~3661--3679, 2015.

\bibitem{Donnell2023}
D.~Donnell, F.~Gao, J.~P. Hughes, B.~Hanscom, L.~Corey, M.~S. Cohen,
  S.~Edupuganti, N.~Mgodi, H.~Rees, J.~M. Baeten, G.~Gray, L.-G. Bekker,
  M.~Hosseinipour, and S.~Delany-Moretlwe, ``Counterfactual estimation of
  efficacy against placebo for novel {PrEP} agents using external trial data:
  example of injectable cabotegravir and oral {PrEP} in women,'' {\em Journal
  of the {I}nternational {AIDS} {S}ociety}, vol.~26, no.~6, p.~e26118, 2023.

\bibitem{Signorovitch2012}
J.~E. Signorovitch, V.~Sikirica, M.~H. Erder, J.~Xie, M.~Lu, P.~S. Hodgkins,
  K.~A. Betts, and E.~Q. Wu, ``Matching-{Adjusted} {Indirect} {Comparisons}:
  {A} {New} {Tool} for {Timely} {Comparative} {Effectiveness} {Research},''
  {\em Value in Health}, vol.~15, pp.~940--947, Sept. 2012.

\bibitem{Gao2021}
F.~Gao, D.~V. Glidden, J.~P. Hughes, and D.~J. Donnell, ``Sample size
  calculation for active-arm trial with counterfactual incidence based on
  recency assay,'' {\em Statistical {C}ommunications in {I}nfectious
  {D}iseases}, vol.~13, no.~1, 2021.

\bibitem{Gao2022}
F.~Gao and M.~Bannick, ``Statistical considerations for cross-sectional {HIV}
  incidence estimation based on recency test,'' {\em Statistics in {M}edicine},
  vol.~41, no.~8, pp.~1446--1461, 2022.

\bibitem{Glidden2020}
D.~V. Glidden, O.~T. Stirrup, and D.~T. Dunn, ``A {B}ayesian averted infection
  framework for {P}r{EP} trials with low numbers of {HIV} infections:
  application to the results of the {DISCOVER} trial,'' {\em The {L}ancet
  {HIV}}, vol.~7, no.~11, pp.~e791--e796, 2020.

\bibitem{Glidden2021}
D.~V. Glidden, M.~Das, D.~T. Dunn, R.~Ebrahimi, Y.~Zhao, O.~T. Stirrup, J.~M.
  Baeten, and P.~L. Anderson, ``Using the adherence-efficacy relationship of
  emtricitabine and tenofovir disoproxil fumarate to calculate background {HIV}
  incidence: a secondary analysis of a randomized, controlled trial,'' {\em
  Journal of the {I}nternational {AIDS} {S}ociety}, vol.~24, no.~5, p.~e25744,
  2021.

\bibitem{Westreich2015}
D.~Westreich and J.~K. Edwards, ``Invited commentary: every good randomization
  deserves observation,'' {\em American {J}ournal of {E}pidemiology}, vol.~182,
  no.~10, pp.~857--860, 2015.

\bibitem{Dahabreh2022}
I.~J. Dahabreh, S.~E. Robertson, and M.~A. Hern{\'a}n, ``Generalizing and
  transporting inferences about the effects of treatment assignment subject to
  non-adherence,'' {\em ar{X}iv preprint ar{X}iv:2211.04876}, 2022.

\bibitem{Keil2014}
A.~P. Keil, J.~K. Edwards, D.~R. Richardson, A.~I. Naimi, and S.~R. Cole, ``The
  parametric {G}-formula for time-to-event data: towards intuition with a
  worked example,'' {\em Epidemiology}, vol.~25, no.~6, p.~889, 2014.

\bibitem{Stuart2011}
E.~A. Stuart, S.~R. Cole, C.~P. Bradshaw, and P.~J. Leaf, ``The use of
  propensity scores to assess the generalizability of results from randomized
  trials,'' {\em Journal of the {R}oyal {S}tatistical {S}ociety: {S}eries {A}
  ({S}tatistics in {S}ociety)}, vol.~174, no.~2, pp.~369--386, 2011.

\bibitem{Dahabreh2020}
I.~J. Dahabreh, L.~C. Petito, S.~E. Robertson, M.~A. Hern{\'a}n, and J.~A.
  Steingrimsson, ``Toward causally interpretable meta-analysis: {T}ransporting
  inferences from multiple randomized trials to a new target population,'' {\em
  Epidemiology}, vol.~31, no.~3, pp.~334--344, 2020.

\bibitem{Harrell1996}
F.~E. Harrell~Jr, K.~L. Lee, and D.~B. Mark, ``Multivariable prognostic models:
  issues in developing models, evaluating assumptions and adequacy, and
  measuring and reducing errors,'' {\em Statistics in {M}edicine}, vol.~15,
  no.~4, pp.~361--387, 1996.

\bibitem{Viele2014}
K.~Viele, S.~Berry, B.~Neuenschwander, B.~Amzal, F.~Chen, N.~Enas, B.~Hobbs,
  J.~G. Ibrahim, N.~Kinnersley, S.~Lindborg, {\em et~al.}, ``Use of historical
  control data for assessing treatment effects in clinical trials,'' {\em
  Pharmaceutical {S}tatistics}, vol.~13, no.~1, pp.~41--54, 2014.

\bibitem{Li2020}
W.~Li, F.~Liu, and D.~Snavely, ``Revisit of test-then-pool methods and some
  practical considerations,'' {\em Pharmaceutical {S}tatistics}, vol.~19,
  no.~5, pp.~498--517, 2020.

\bibitem{Carr1998}
A.~Carr, K.~Samaras, D.~J. Chisholm, and D.~A. Cooper, ``Pathogenesis of
  {HIV}-1-protease inhibitor-associated peripheral lipodystrophy,
  hyperlipidaemia, and insulin resistance,'' {\em The {L}ancet}, vol.~351,
  no.~9119, pp.~1881--1883, 1998.

\bibitem{ShookSa2023}
B.~E. Shook-Sa, P.~N. Zivich, S.~P. Rosin, J.~K. Edwards, A.~A. Adimora, M.~G.
  Hudgens, and S.~R. Cole, ``Fusing {Trial} {Data} for {Treatment}
  {Comparisons}: {Single} versus {Multi}-{Span} {Bridging},'' May 2023.
\newblock arXiv:2305.00845 [stat].

\bibitem{Zivich2023p}
P.~N. Zivich, S.~R. Cole, J.~K. Edwards, G.~E. Mulholland, B.~E. Shook-Sa, and
  E.~J. Tchetgen~Tchetgen, ``{INTRODUCING} {PROXIMAL} {CAUSAL} {INFERENCE}
  {FOR} {EPIDEMIOLOGISTS},'' {\em American Journal of Epidemiology}, vol.~192,
  pp.~1224--1227, July 2023.

\bibitem{Daniel2014}
R.~M. Daniel, ``Double robustness,'' in {\em Wiley {S}tats{R}ef: {S}tatistics
  {R}eference {O}nline}, John Wiley \& Sons, Ltd, 2014.

\bibitem{Rotnitsky2006}
A.~Rotnitzky and J.~M. Robins, ``Inverse {P}robability {W}eighting in
  {S}urvival {A}nalysis,'' in {\em Wiley {S}tats{R}ef: {S}tatistics {R}eference
  {O}nline}, John Wiley \& Sons, Ltd, 2014.

\bibitem{Zivich2021a}
P.~N. Zivich and A.~Breskin, ``Machine learning for causal inference: on the
  use of cross-fit estimators,'' {\em Epidemiology}, vol.~32, no.~3,
  pp.~393--401, 2021.

\bibitem{Naimi2017}
A.~I. Naimi, A.~E. Mishler, and E.~H. Kennedy, ``Challenges in obtaining valid
  causal effect estimates with machine learning algorithms,'' {\em American
  {J}ournal of {E}pidemiology}, 07 2021.
\newblock kwab201.

\bibitem{Lee2008}
S.-H. Lee, E.-J. Lee, and B.~O. Omolo, ``Using integrated weighted survival
  difference for the two-sample censored data problem,'' {\em Computational
  Statistics \& Data Analysis}, vol.~52, pp.~4410--4416, May 2008.

\bibitem{Fleming1980}
T.~R. Fleming, J.~R. O'Fallon, P.~C. O'Brien, and D.~P. Harrington, ``Modified
  {Kolmogorov}-{Smirnov} {Test} {Procedures} with {Application} to
  {Arbitrarily} {Right}-{Censored} {Data},'' {\em Biometrics}, vol.~36, no.~4,
  pp.~607--625, 1980.
\newblock Publisher: [Wiley, International Biometric Society].

\bibitem{Harris2020}
C.~R. Harris, K.~J. Millman, S.~J. Van Der~Walt, R.~Gommers, P.~Virtanen,
  D.~Cournapeau, E.~Wieser, J.~Taylor, S.~Berg, N.~J. Smith, {\em et~al.},
  ``Array programming with {N}um{P}y,'' {\em Nature}, vol.~585, no.~7825,
  pp.~357--362, 2020.

\bibitem{Virtanen2020}
P.~Virtanen, R.~Gommers, T.~E. Oliphant, M.~Haberland, T.~Reddy, D.~Cournapeau,
  E.~Burovski, P.~Peterson, W.~Weckesser, J.~Bright, {\em et~al.}, ``Sci{P}y
  1.0: fundamental algorithms for scientific computing in {P}ython,'' {\em
  Nature {M}ethods}, vol.~17, no.~3, pp.~261--272, 2020.

\bibitem{Mckinney2010}
{W}es {M}c{K}inney, ``{D}ata {S}tructures for {S}tatistical {C}omputing in
  {P}ython,'' in {\em {P}roceedings of the 9th {P}ython in {S}cience
  {C}onference} ({S}t\'efan van~der {W}alt and {J}arrod {M}illman, eds.),
  pp.~56 -- 61, 2010.

\bibitem{Seabold2010}
S.~Seabold and J.~Perktold, ``statsmodels: {E}conometric and statistical
  modeling with {P}ython,'' in {\em 9th {P}ython in {S}cience {C}onference},
  2010.

\bibitem{Hunter2007}
J.~D. Hunter, ``Matplotlib: A 2{D} graphics environment,'' {\em Computing in
  {S}cience \& {E}ngineering}, vol.~9, no.~3, pp.~90--95, 2007.

\end{thebibliography}
\bibliographystyle{ieeetr}

\newpage 

\begin{table}[h]
	\caption{Descriptive statistics for baseline characteristics of participants in the AIDS Clinical Trial Group 175 and 320 randomized trials by assigned antiretroviral therapy}
	\centering
	\begin{tabular}{llccccc}
		\hline
		&                          & \multicolumn{2}{c}{ACTG 175\textsuperscript{*}}                                                                                                   &  & \multicolumn{2}{c}{ACTG 320\textsuperscript{†}}                                                                                                       \\
		&                          & \begin{tabular}[c]{@{}c@{}}Mono therapy\\ (n=271)\end{tabular} & \begin{tabular}[c]{@{}c@{}}Dual therapy\\ (N=542)\end{tabular} &  & \begin{tabular}[c]{@{}c@{}}Dual therapy \\ (n=579)\end{tabular} & \begin{tabular}[c]{@{}c@{}}Triple therapy\\ (n=577)\end{tabular} \\ \cline{3-4} \cline{6-7} 
		Male                &                          & 221 (82\%)                                                    & 440 (81\%)                                                     &  & 485 (84\%)                                                      & 471 (82\%)                                                       \\
		\multicolumn{2}{l}{Age {[}Q1, Q3{]}}              & 35 {[}30, 41{]}                                               & 35 {[}30, 41{]}                                                &  & 38 {[}33, 44{]}                                                 & 38 {[}33, 44{]}                                                  \\
		\multicolumn{2}{l}{Black}                      & 70 (26\%)                                                     & 138 (26\%)                                                     &  & 165 (28\%)                                                      & 163 (28\%)                                                       \\
		\multicolumn{2}{l}{History of IDU}             & 33 (12\%)                                                     & 79 (15\%)                                                      &  & 93 (16\%)                                                       & 91 (16\%)                                                        \\
		\multicolumn{2}{l}{Karnofsky Score {[}Q1, Q3{]}}  & 100 {[}90, 100{]}                                             & 100 {[}90, 100{]}                                              &  & 90 {[}90, 100{]}                                                & 90 {[}90, 100{]}                                                 \\
		\multicolumn{2}{l}{Karnofsky Score Categories} &                                                               &                                                                &  &                                                                 &                                                                  \\
		& 100                      & 153 (56\%)                                                    & 317 (58\%)                                                     &  & 202 (35\%)                                                      & 195 (34\%)                                                       \\
		& 90                       & 102 (38\%)                                                    & 200 (37\%)                                                     &  & 269 (46\%)                                                      & 276 (48\%)                                                       \\
		& \textless{}90            & 16 (6\%)                                                      & 25 (5\%)                                                       &  & 108 (19\%)                                                      & 106 (18\%)                                                       \\
		\multicolumn{2}{l}{CD4 {[}Q1, Q3{]}}              & 321 {[}256, 415{]}                                            & 323 {[}247, 416{]}                                             &  & 70 {[}22, 135{]}                                                & 80 {[}24, 138{]}                                                 \\ \hline
	\end{tabular}
	\floatfoot{ACTG: AIDS Clinical Trial Group, Q1: first quartile, Q3: third quartile, IDU: injection drug use.\\
	Trials were harmonized by restricting the ACTG 175 to participants with at least three months of prior zidovudine treatment and no prior non-zidovudine antiretroviral therapy (821 of 2493 participants), restricting to participants at least 16 years of age (813 of 821), administratively censoring participants in ACTG 175 at one year, and using the ACTG 175 composite outcome of AIDS, death, or a 50\%+ decline in CD4 cell count for both trials.\\
	* For ACTG 175, dual therapy consisted of either zidovudine-didanosine or zidovudine-zalcitabine. Mono therapy consisted of zidovudine-only.\\
	† For ACTG 320, dual therapy consisted of zidovudine-lamivudine. Triple therapy consisted of zidovudine-lamivudine-indinavir.}
\end{table}

\begin{table}[H]
	\caption{Descriptive statistics for baseline characteristics of participants in the AIDS Clinical Trial Group 175 and 320 randomized trials by assigned antiretroviral therapy and restricted by baseline CD4 counts}
	\centering
	\begin{tabular}{llccccc}
		\hline
		&                          & \multicolumn{2}{c}{ACTG 175}                                                                                                   &  & \multicolumn{2}{c}{ACTG 320}                                                                                                       \\
		&                          & \begin{tabular}[c]{@{}c@{}}Mono therapy\\ (n=110)\end{tabular} & \begin{tabular}[c]{@{}c@{}}Dual therapy\\ (N=224)\end{tabular} &  & \begin{tabular}[c]{@{}c@{}}Dual therapy \\ (n=344)\end{tabular} & \begin{tabular}[c]{@{}c@{}}Triple therapy\\ (n=356)\end{tabular} \\ \cline{3-4} \cline{6-7} 
		Male                &                          & 93 (85\%)                                                     & 186 (83\%)                                                     &  & 285 (83\%)                                                      & 284 (80\%)                                                       \\
		\multicolumn{2}{l}{Age {[}Q1, Q3{]}}              & 35 {[}30, 41{]}                                               & 35 {[}30, 41{]}                                                &  & 38 {[}33, 44{]}                                                 & 38 {[}33, 45{]}                                                  \\
		\multicolumn{2}{l}{Black}                      & 35 (32\%)                                                     & 52 (23\%)                                                      &  & 97 (28\%)                                                       & 89 (25\%)                                                        \\
		\multicolumn{2}{l}{History of IDU}             & 13 (12\%)                                                     & 27 (12\%)                                                      &  & 56 (16\%)                                                       & 56 (16\%)                                                        \\
		\multicolumn{2}{l}{Karnofsky Score {[}Q1, Q3{]}}  & 100 {[}90, 100{]}                                             & 100 {[}90, 100{]}                                              &  & 90 {[}90, 100{]}                                                & 90 {[}90, 100{]}                                                 \\
		\multicolumn{2}{l}{Karnofsky Score Categories} &                                                               &                                                                &  &                                                                 &                                                                  \\
		& 100                      & 60 (55\%)                                                     & 122 (54\%)                                                     &  & 166 (48\%)                                                      & 167 (47\%)                                                       \\
		& 90                       & 40 (36\%)                                                     & 84 (37\%)                                                      &  & 136 (39\%)                                                      & 139 (39\%)                                                       \\
		& \textless{}90            & 10 (9\%)                                                      & 18 (8\%)                                                       &  & 42 (12\%)                                                       & 50 (14\%)                                                        \\
		\multicolumn{2}{l}{CD4 {[}Q1, Q3{]}}              & 240 {[}210, 275{]}                                            & 236 {[}207, 267{]}                                             &  & 118 {[}80, 158{]}                                               & 124 {[}88, 165{]}                                                \\ \hline
	\end{tabular}
	\floatfoot{ACTG: AIDS Clinical Trial Group, Q1: first quartile, Q3: third quartile, IDU: injection drug use.\\
	Trials were harmonized by restricting the ACTG 175 to participants with at least three months of prior zidovudine treatment and no prior non-zidovudine antiretroviral therapy (821 of 2493 participants), restricting to participants at least 16 years of age (813 of 821), administratively censoring participants in ACTG 175 at one year, and using the ACTG 175 composite outcome of AIDS, death, or a 50\%+ decline in CD4 cell count for both trials. The population was further restricted by baseline CD4 cell counts between 50-300 cells/mm\textsuperscript{3}\\
	* For ACTG 175, dual therapy consisted of either zidovudine-didanosine or zidovudine-zalcitabine. Mono therapy consisted of zidovudine-only.\\
	† For ACTG 320, dual therapy consisted of zidovudine-lamivudine. Triple therapy consisted of zidovudine-lamivudine-indinavir.}
\end{table}

\begin{figure}[H]
	\centering
	\caption{Twister plots for difference between the dual antiretroviral therapy arms of the ACTG 320 and ACTG 175 trials}
	\includegraphics[width=1.0\linewidth]{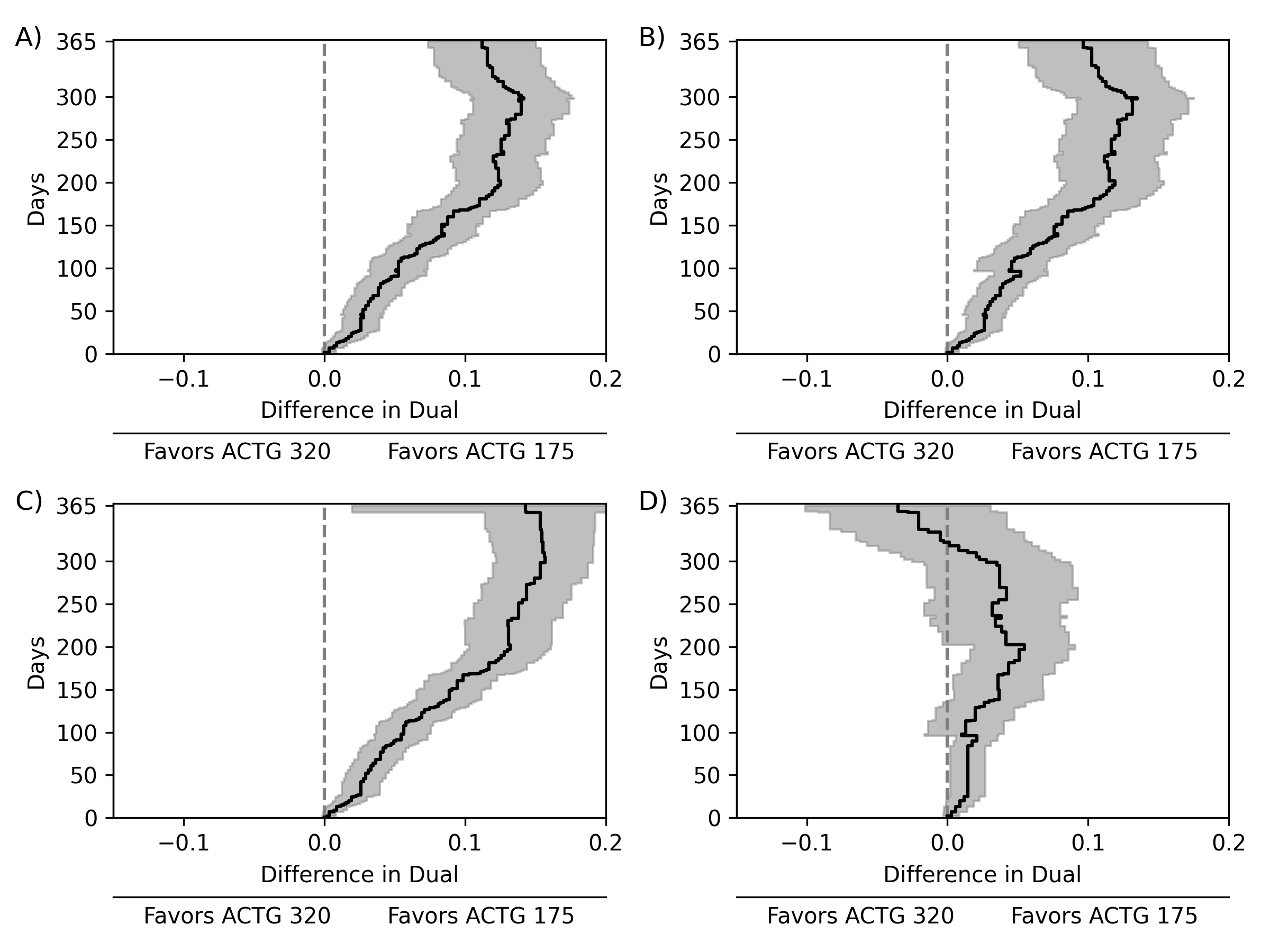}
	\floatfoot{A: pre-standardization, B: standardized to the ACTG 320 population, C: added CD4 to the inverse odds of sampling weight model, D: restricted target population by CD4 (baseline CD4 between 50 and 300 cells/mm\textsuperscript{3}). Shaded regions indicated 95\% Wald-type confidence intervals.}
\end{figure}

\begin{figure}[H]
	\centering
	\caption{Estimated risk difference comparing triple therapy versus mono therapy for the CD4-restricted ACTG 320 population}
	\includegraphics[scale=0.9]{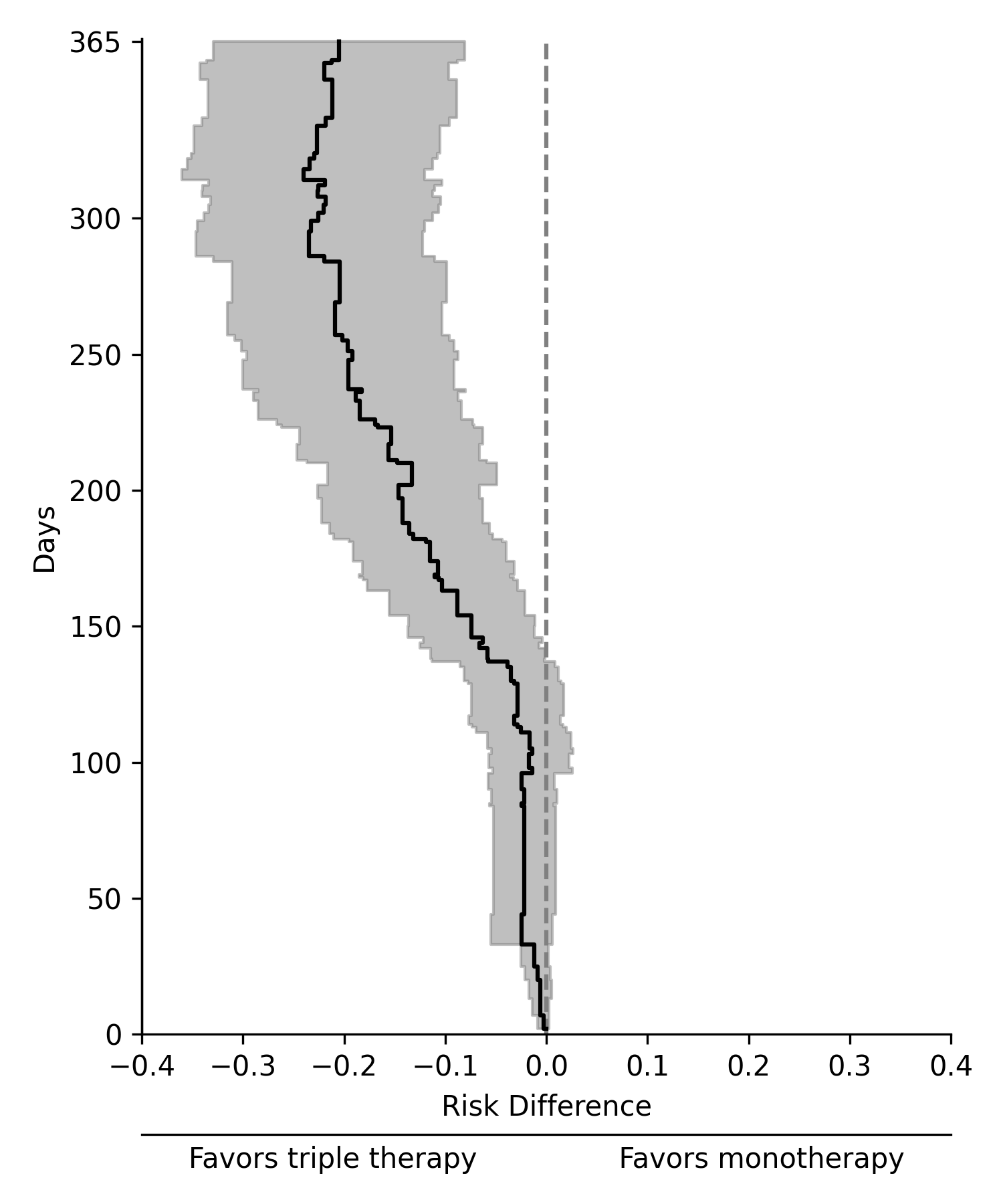}
	\floatfoot{Shaded regions indicated 95\% Wald-type confidence intervals. Target population restricted to CD4 cell counts between 50-300 cells/mm\textsuperscript{3} at baseline.}
\end{figure}

\newpage

\section*{Appendix}

\renewcommand{\thefigure}{A\arabic{figure}}
\setcounter{figure}{0}

\begin{figure}[H]
	\centering
	\caption{Distribution of baseline CD4 cell counts by AIDS Clinical Trial Group}
	\includegraphics[scale=0.9]{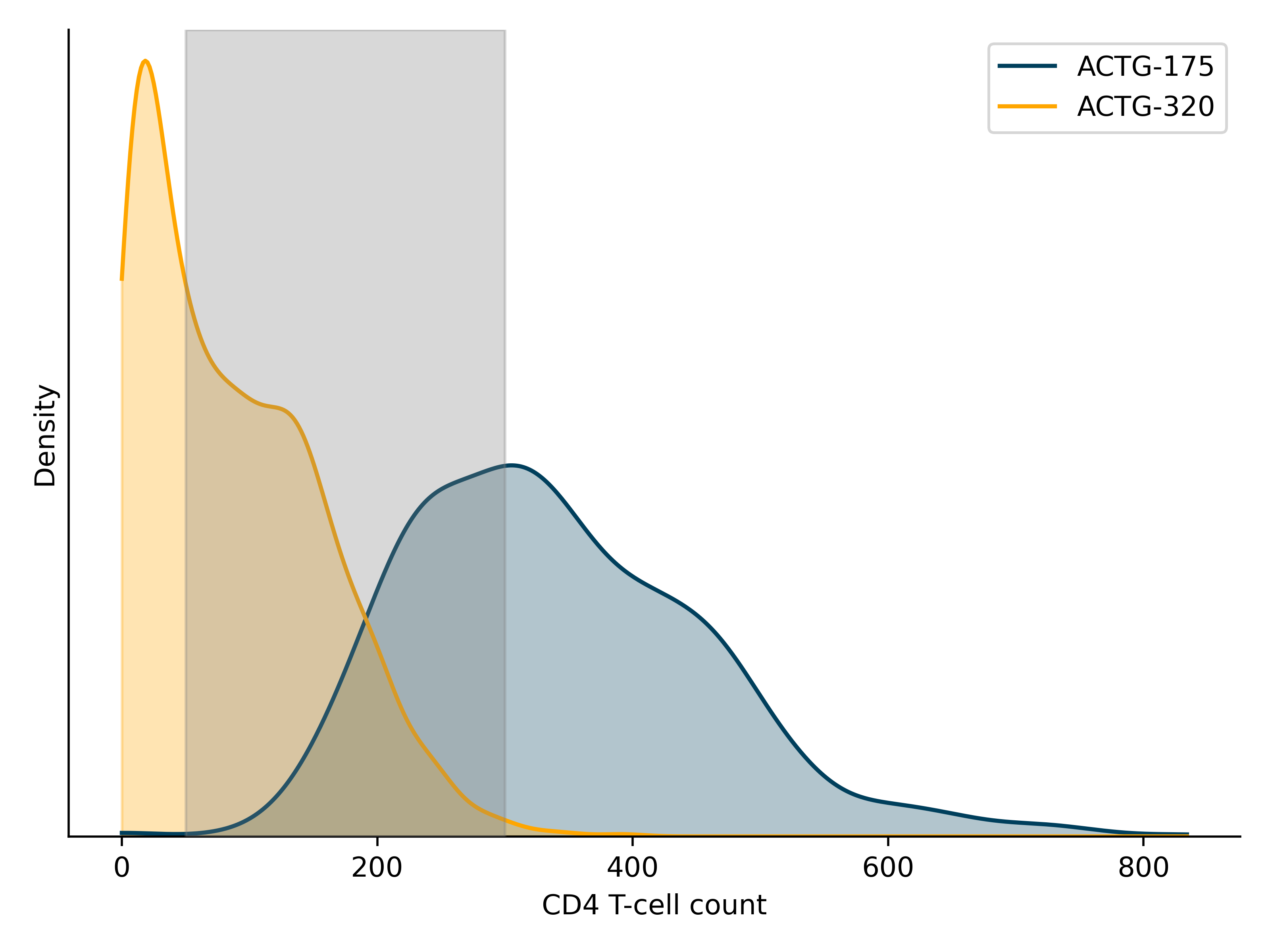}
	\floatfoot{ACTG: AIDS Clinical Trial Group.\\
		Plots were created using a Gaussian kernel density estimator stratified by trial. The gray shaded region indicates baseline CD4 between 50-300 cells/mm\textsuperscript{3}.}
\end{figure}

\newpage 

\subsection*{Appendix 1: Proposed fusion diagnostic}

As a statistical diagnostic, we propose comparing the difference between the estimated risk functions of the shared intermediate arms from the two trials. As implied by the identification assumptions, the risk functions for the shared arms in the target population should be the same. This suggests using a test statistic such as
\[\hat{\mu} = \int_{0}^{\tau} \left\{ \hat{R}^2_{320}(t) - \hat{R}^2_{175}(t) \right\} dt\]
which compares the risk function for $A=2$ estimated from each trial. A related test statistic is the weighted Kaplan-Meier statistic of Pepe and Fleming (1989) \cite{Pepe1989}, defined as
\[\hat{\mu}_{WKM} = \int_{0}^{\tau} \hat{w}(t) \left\{ \hat{S}^1(t) - \hat{S}^0(t) \right\} dt\]
where $S^a(t)$ is a stratified Kaplan-Meier for $A=a$ and $\hat{w}(t)$ is a user-specified weight function. The purpose of the weight function is to improve stability of the weighted Kaplan-Meier statistic in cases of heavy censoring by down-weighting differences at times later in follow-up. Pepe and Fleming showed the weighted Kaplan-Meier test statistic is asymptotically normal \cite{Pepe1989, Pepe1991}, which can be used to approximate the sampling distribution of the statistic under the null hypothesis that the two survival curves are equal. Unlike tests based on the stochastic ordering (e.g., log-rank), measures based on the integrated difference may be sensitive to differences in survival functions even in cases where the hazard functions cross \cite{Pepe1989, Pepe1991, Lee2008}. Further, the weighted Kaplan-Meier statistic can be sensitive to moderate differences over extended periods of time, unlike modifications of the Kolmogorov-Smirnov test statistic which are based on the maximum distance between the survival functions \cite{Fleming1980}. 

The proposed diagnostic statistic $\hat{\mu}$ can be viewed as a modification of the weighted Kaplan-Meier statistic, where survival is replaced with risk, the Kaplan-Meier estimator is replaced with the inverse probability weighted empirical distribution function estimator \cite{},
and $\hat{w}(t) = 1$ for all $t$. To determine whether $\hat{\mu}$ is further from zero than would be expected by chance, the standard error of $\hat{\mu}$ is estimated using a nonparametric bootstrap. The bootstrap can be implemented using the following algorithm:

\begin{algorithm}[H]
	\caption{Diagnostic test}\label{mestr}
	\begin{algorithmic}%[1]
		\State \textbf{calculate} the estimated risk functions, $\widehat{R}_{320}^{2}(t)$ and $\widehat{R}_{175}^{2}(t)$
		\State \textbf{calculate} $\hat{\mu}$ (see algorithm below)
		\For{$j \gets 1, b$}
			\State \textbf{randomly sample} with replacement $n_1$ observations from those with $S=1$
			\State \textbf{randomly sample} with replacement $n_0$ observations from those with $S=0$
			\State \textbf{estimate} the nuisance models using the resampled data
			\State \textbf{estimate} the risk functions $\widehat{R^*}_{320}^{2}(t)$ and $\widehat{R^*}_{175}^{2}(t)$ using the resampled data
			
			\State \textbf{calculate} $\hat{\mu}^*$ using $\widehat{R^*}_{320}^{2}(t)$ and $\widehat{R^*}_{175}^{2}(t)$
			\State \textbf{save} $\hat{\mu}^*$
		\EndFor
		\State \textbf{estimate} the standard error of $\hat{\mu}$ via the standard deviation of the $b$ balues of $\hat{\mu}^*$
		\State \textbf{calculate} the Z-score using $\frac{\hat{\mu}}{\widehat{SE}(\hat{\mu})}$
		\\
		
		\Procedure{Integrated risk difference}{}
			\State (1) Align the risk functions by the union of the unique events times
			\State (2) Subtract the next event time from the current event time
			\State (3) Subtract estimated risks of shared arms at each unique event time
			\State (4) Multiply the time differences (2) by the estimated risk differences in (3)
			\State (5) Sum the multiplied differences from (4)
		\EndProcedure
		\\
	\end{algorithmic}
\end{algorithm}

\newpage
An illustration of the calculation of the area between the risk functions. Note that the values of $t$ are the union of unique event times from each of the data sources
\begin{table}[H]
	\begin{tabular}{lcccccc}
		\hline
		k & \multicolumn{3}{c}{(1)}                                         & (2)             & (3)                       & (4)                                                \\
		& $t$ & $\widehat{R}_{320}^{2}(t_k)$ & $\widehat{R}_{175}^2(t_k)$ & $t_{k+1} - t_k$ & $\widehat{RD}^{2-2}(t_k)$ & $|\widehat{RD}^{2-2}(t_k)| \times (t_{k+1} - t_k)$ \\ \cline{2-7} 
		1 & 0   & 0.00                         & 0.00                       & 0.2-0=0.2       & 0                         & $0 \times 0.2 = 0$                               \\
		2 & 0.2 & 0.00                         & 0.07                       & 0.4-0.2=0.2     & -0.07                     & $-0.07 \times 0.2 = -0.014$                       \\
		3 & 0.4 & 0.10                         & 0.07                       & 1.2-0.4=0.8     & 0.03                      & $0.03 \times 0.8 = 0.024$                        \\
		4 & 1.2 & 0.20                         & 0.07                       & 1.7-1.2=0.5     & 0.13                      & $0.13 \times 0.5 = 0.065$                        \\
		5 & 1.7 & 0.35                         & 0.30                       & 2.4-1.7=0.7     & 0.05                      & $0.05 \times 0.7 = 0.035$                        \\
		6 & 2.4 & 0.35                         & 0.45                       & 2.5-2.5=0.1     & -0.10                     & $-0.10 \times 0.1 = -0.01$                        \\
		7 & 2.5 & 0.45                         & 0.45                       & 3.0-2.5=0.5     & 0.00                      & $0.00 \times 0.5 = 0$                            \\
		8 & 3.0 & 0.45                         & 0.55                       & 0               & -0.10                     & $-0.10 \times 0 = 0$                             \\ \hline
	\end{tabular}
\end{table}
(5) $\sum_{k=1}^{8} |\widehat{RD}^{2-2}(t_k)| \times (t_{k+1} - t_k) = 0.100$

\newpage 

\subsection*{Appendix 2: Simulation study}

\renewcommand{\thetable}{A\arabic{figure}}
\setcounter{table}{0}

To assess the type 1 error and power of the proposed diagnostic test, a simulation study was conducted mimicking the ACTG example with monotherapy ($A=1$), dual therapy ($A=2$), and triple therapy ($A=3$). The parameter of interest compared triple therapy to mono therapy. Let $n_1 = \sum_{i=1}^n I(S_i = 1)$ and $n_0 = \sum_{i=1}^n I(S_i = 0)$. Observations were generated as follows:
\[W_{1,i} = \text{Bernoulli}(0.3)\]
\[Z_{i} = \text{Normal}(\mu=0.3,\sigma=45)\]
\[W_{2,i} = 
\begin{cases}
	0 & \text{if } -7 W_{1,i} + Z_i < 0\\    
	-7 W_{1,i} + Z_i & \text{if } 0 \le -7 W_{1,i} + Z_i \le 1600\\    
	1600 & \text{if } -7 W_{1,i} + Z_i > 1600
\end{cases}\]
\[S_i = \text{Bernoulli}(\text{expit}(0.5 - 1.5W_{1,i} + 0.02(W_{2,i} - 250)))\]
\[A_i =
\begin{cases}
	\text{Bernoulli}(0.5) & \text{if } S_i=0\\    
	\text{Bernoulli}(0.5) + 1 & \text{if } S_i = 1
\end{cases}\]
where $W_{1}$ is history of injection drug use and $W_2$ is baseline CD4. Potential event times were simulated from a Weibull distribution. In particular,
\[T_i^a = \left(-\lambda(a,W_{1,i},W_{2,i}) \ln(U_i) \right)^{0.8}\]
where $U_i$ is a random draw from a uniform distribution over $(0,1)$, and
\[\lambda(a,W_{1,i},W_{2,i}) = \exp\left(4.9 + 0.4\times I(a=2) + 1.5 \times I(a=3) - 3W_{1,i} + 0.01 W_{2,i} - 0.2 I(a=2)W_{1,i} - 0.25 I(a=3)W_{1,i}\right)\]
The event time under the observed treatment, $T_i$, was
\[T_i = \sum_{a\in\{1,2,3\}} I(A_i=a) T_i^a\]
Censoring times were generated from the following Weibull model
\[C_i = \left(-7.2 \ln(\tilde{U}_i)\right)^3\]
where $\tilde{U}_i$ is a random draw from a uniform distribution over $(0,1)$. All observations were administratively censored at $t=365$. Therefore, the observed time and event indicator were 
\[T_i^* = \text{min}(T_i, C_i, 365)\]
\[\delta_i = I(T_i = T_i^*)\]
To generate the trial data sets, $3\times(n_1+n_0)$ observations were first generated as described above. Then $n_1$ and $n_0$ observations were randomly selected among those with $S_i=1$ and $S_i=0$, respectively.

Two variations of the fusion estimator were compared based on an incorrectly specified sampling model and correctly specified sampling model. The incorrectly specified fusion estimator used the following sampling model
\[\pi_S(W_i,\beta) = \beta_0 + \beta_1 W_{2,i}\]
whereas the correctly specified fusion estimator used the following sampling model specification
\[\pi_S(W_i,\beta) = \beta_0 + \beta_1 W_{1,i} + \beta_2 W_{2,i}\]
The probability of remaining uncensored was estimated via the Nelson-Aalen estimator since the censoring mechanism did not depend on any covariates.

We simulated two different sample sizes: ($n_1 = n_0 = 1000$) and ($n_1 = n_0 = 2000$). For each of the sample sizes, 2000 simulated data sets were generated. Type 1 error was estimated by the proportion of P values below a selected significance level under the correctly specified sampling model. Power estimated by the proportion of P values below a selected significance level under the incorrectly specified model. Significance levels of 0.05, 0.10, and 0.20 were considered.  In addition to assessment of the proposed diagnostic, we also assessed bias, empirical standard error, standard error ratio, and 95\% Wald confidence interval coverage at $t\in\{91, 183, 274, 365\}$ for the risk difference \cite{Morris2022}. Bias was defined as the mean of the difference between the estimated risk difference each simulated data set and the truth. The true risk difference was estimated from the potential outcomes $T^{a=3}$ and $T^{a=1}$ for 20,000,000 generated observations. Empirical standard error was estimated by the standard deviation of the point estimates of the simulation. The standard error ratio was defined as the empirical standard error divided by the mean of the estimated standard error across the data sets. Confidence interval coverage was defined as the proportion of confidence intervals that contained the true risk difference. The standard error estimates were based on a nonparametric bootstrap with 500 data sets resampled with replacement. The correct sampling model results were further examined in the subset of simulated data sets where the permutation P value was above the significance threshold. All simulations were conducted using Python 3.6.8, with the following libraries: \texttt{NumPy} \cite{Harris2020}, \texttt{SciPy} \cite{Virtanen2020}, \texttt{pandas} \cite{Mckinney2010}, \texttt{statsmodels} \cite{Seabold2010}, and \texttt{matplotlib} \cite{Hunter2007}. For the diagnostic test, type 1 error was near the nominal level (Appendix Table A1). Power was 1 for bias that varied from about 0.014 to 0.037 over follow-up, meaning that incorrectly specified sampling models were always detected. Notably, this bias was small relative to the standard error of the estimator. When the sampling model was correctly specified, there was little bias fro the risk difference (Appendix Tables A2-A3). As $\alpha$ increased for the diagnostic test, confidence interval coverage was increasingly conservative. When the sampling model was incorrectly specified, bias was present across selected time points and sample sizes.

\begin{table}[H]
	\caption{Estimated type 1 error and power of the proposed diagnostic permutation test}
	\centering
	\begin{tabular}{lcccccc}
		\hline
		& \multicolumn{2}{c}{$\alpha = 0.05$} & \multicolumn{2}{c}{$\alpha=0.10$} & \multicolumn{2}{c}{$\alpha=0.20$} \\
		& Type 1            & Power           & Type 1           & Power          & Type 1           & Power          \\ \cline{2-7} 
		$n_1 = n_0 = 1000$ & 0.058              & 1.000            & 0.109             & 1.000           & 0.209             & 1.000           \\
		$n_1 = n_0 = 2000$ & 0.057              & 1.000            & 0.113             & 1.000           & 0.219             & 1.000           \\ \hline
	\end{tabular}
	\floatfoot{Results are reported for 2000 generated data sets. Type 1 error was estimated by the proportion of \textit{P} values below the corresponding $\alpha$ for the correctly specified sampling model. Power was estimated by the proportion of \textit{P} values below the corresponding $\alpha$ for the incorrectly specified sampling model}
\end{table}

\newpage 

\renewcommand{\thetable}{A2}

\begin{table}[H]
	\caption{Simulation study results for the fusion estimators with $n_1 = n_0 = 1000$}
	\centering
	\begin{tabular}{llcccc}
		\hline
		&                            & Bias   & ESE   & SER  & 95\% CI Coverage \\ \cline{3-6} 
		\multicolumn{2}{l}{Day 91}    &        &       &      &                  \\
		& Incorrect sampling model   & 0.014  & 0.058 & 1.01 & 94\%             \\
		& Correct sampling model     & -0.001 & 0.046 & 1.00 & 95\%             \\
		& Correct with $\alpha=0.05$ &  0.000 & 0.044 & 1.04 & 96\%             \\
		& Correct with $\alpha=0.10$ & -0.001 & 0.043 & 1.08 & 97\%             \\
		& Correct with $\alpha=0.20$ & -0.001 & 0.042 & 1.10 & 97\%             \\
		\multicolumn{2}{l}{Day 183}   &        &       &      &                  \\
		& Incorrect sampling model   & 0.028  & 0.062 & 1.00 & 93\%             \\
		& Correct sampling model     & 0.001  & 0.056 & 0.97 & 95\%             \\
		& Correct with $\alpha=0.05$ & 0.001  & 0.053 & 1.03 & 96\%             \\
		& Correct with $\alpha=0.10$ & 0.001  & 0.051 & 1.07 & 97\%             \\
		& Correct with $\alpha=0.20$ & 0.000  & 0.049 & 1.11 & 97\%             \\
		\multicolumn{2}{l}{Day 274}   &        &       &      &                  \\
		& Incorrect sampling model   & 0.033  & 0.063 & 1.00 & 91\%             \\
		& Correct sampling model     & -0.001 & 0.063 & 0.97 & 94\%             \\
		& Correct with $\alpha=0.05$ & -0.001 & 0.059 & 1.02 & 95\%             \\
		& Correct with $\alpha=0.10$ &  0.000 & 0.057 & 1.07 & 96\%             \\
		& Correct with $\alpha=0.20$ &  0.000 & 0.055 & 1.11 & 97\%             \\
		\multicolumn{2}{l}{Day 365}   &        &       &      &                  \\
		& Incorrect sampling model   & 0.037 & 0.068 & 0.98 & 91\%             \\
		& Correct sampling model     & 0.000 & 0.071 & 0.96 & 94\%             \\
		& Correct with $\alpha=0.05$ & 0.001 & 0.068 & 1.01 & 95\%             \\
		& Correct with $\alpha=0.10$ & 0.001 & 0.066 & 1.04 & 96\%             \\
		& Correct with $\alpha=0.20$ & 0.001 & 0.064 & 1.08 & 97\%             \\ \cline{2-6} 
	\end{tabular}
	\floatfoot{ESE: empirical standard error of the estimator, SER: standard error ratio, CI: confidence interval. Bias was defined as the mean of the estimated risk difference at the corresponding time minus the true risk difference. ESE: was defied as the standard deviation of the simulation estimates at the corresponding time. SER was the mean of the estimated standard errors across all simulations divided by the ESE. 95\% CI coverage was defined as the proportion of 95\% CIs containing the true risk difference. $\alpha$ was the cut-point used for the diagnostic procedure; when the corresponding P value was below the $\alpha$ threshold, those simulation iterations were dropped. \\
	The incorrectly specified sampling model included only CD4. The correctly specified sampling model included CD4 and injection drug use.}
\end{table}

\renewcommand{\thetable}{A3}

\begin{table}[H]
	\caption{Simulation study results for the fusion estimators with $n_1 = n_0 = 2000$}
	\begin{tabular}{llcccc}
		\hline
		&                            & Bias   & ESE   & SER  & 95\% CI Coverage \\ \cline{3-6} 
		\multicolumn{2}{l}{Day 91}    &        &       &      &                  \\
		& Incorrect sampling model   & 0.014 & 0.041 & 1.02 & 95\%             \\
		& Correct sampling model     & 0.000 & 0.032 & 1.02 & 95\%             \\
		& Correct with $\alpha=0.05$ & 0.001 & 0.031 & 1.07 & 96\%             \\
		& Correct with $\alpha=0.10$ & 0.001 & 0.030 & 1.10 & 97\%             \\
		& Correct with $\alpha=0.20$ & 0.001 & 0.029 & 1.14 & 97\%             \\
		\multicolumn{2}{l}{Day 183}   &        &       &      &                  \\
		& Incorrect sampling model   & 0.027 & 0.043 & 1.01 & 91\%             \\
		& Correct sampling model     & 0.000 & 0.038 & 1.01 & 95\%             \\
		& Correct with $\alpha=0.05$ & 0.001 & 0.036 & 1.07 & 96\%             \\
		& Correct with $\alpha=0.10$ & 0.001 & 0.035 & 1.11 & 96\%             \\
		& Correct with $\alpha=0.20$ & 0.001 & 0.033 & 1.16 & 97\%             \\
		\multicolumn{2}{l}{Day 274}   &        &       &      &                  \\
		& Incorrect sampling model   & 0.033  & 0.045 & 1.02 & 89\%             \\
		& Correct sampling model     & 0.000  & 0.043 & 1.00 & 95\%             \\
		& Correct with $\alpha=0.05$ & 0.001  & 0.041 & 1.06 & 96\%             \\
		& Correct with $\alpha=0.10$ & 0.001  & 0.040 & 1.09 & 97\%             \\
		& Correct with $\alpha=0.20$ & 0.001  & 0.037 & 1.15 & 98\%             \\
		\multicolumn{2}{l}{Day 365}   &        &       &      &                  \\
		& Incorrect sampling model   & 0.037  & 0.048 & 1.00 & 87\%             \\
		& Correct sampling model     & 0.001  & 0.049 & 0.99 & 94\%             \\
		& Correct with $\alpha=0.05$ & 0.002  & 0.047 & 1.03 & 95\%             \\
		& Correct with $\alpha=0.10$ & 0.003  & 0.046 & 1.05 & 96\%             \\
		& Correct with $\alpha=0.20$ & 0.002  & 0.044 & 1.11 & 97\%             \\ \cline{2-6} 
	\end{tabular}
	\floatfoot{ESE: empirical standard error of the estimator, SER: standard error ratio, CI: confidence interval. Bias was defined as the mean of the estimated risk difference at the corresponding time minus the true risk difference. ESE: was defied as the standard deviation of the simulation estimates at the corresponding time. SER was the mean of the estimated standard errors across all simulations divided by the ESE. 95\% CI coverage was defined as the proportion of 95\% CIs containing the true risk difference. $\alpha$ was the cut-point used for the diagnostic procedure; when the corresponding P value was below the $\alpha$ threshold, those simulation iterations were dropped. \\
	The incorrectly specified sampling model included only CD4. The correctly specified sampling model included CD4 and injection drug use.}
\end{table}

\end{document}